\documentclass{article}
\usepackage{graphicx}
\graphicspath{ {./images/} }
\usepackage[T1]{fontenc}
\usepackage[english]{babel}
\usepackage{prettyref}
\usepackage[autostyle]{csquotes}
\usepackage{indentfirst}
\usepackage{microtype}
\usepackage[
    unicode=true,
    bookmarks=true, bookmarksnumbered=true, bookmarksopen=true, bookmarksopenlevel=1,
    breaklinks=true, pdfborder={0 0 0}, pdfborderstyle={}, backref=false, colorlinks=false,
]{hyperref}
\usepackage{graphicx}
\usepackage{listings}
\usepackage{mathtools,amstext,amsmath,amssymb,amsfonts}
\usepackage{enumitem}
\usepackage{tabularx}
\usepackage{ragged2e}
\newcolumntype{Y}{>{\RaggedRight\arraybackslash}X}
\usepackage{booktabs}

\graphicspath{ {./Images/} }

\newcommand{\ket}[1]{\left\vert #1\right\rangle}
\DeclareMathOperator*{\argmin}{argmin}

\hypersetup{
    pdftitle={Software Verification using Quantum Computers},
    pdfauthor={Sebastian Issel},
    pdfpagelayout=OneColumn, pdfnewwindow=true, pdfstartview=XYZ, plainpages=false,
}

\usepackage{biblatex}
\begin{document}
\title{Towards Classical Software Verification using Quantum Computers}
\author{
    Sebastian~Issel, Kilian~Tscharke and Pascal~Debus\\
    \textit{<Firstname>}.\textit{<Lastname>}@AISEC.Fraunhofer.de\\
    Fraunhofer AISEC\\
    Lichtenbergstr. 11\\
    85748 Garching Germany
}
\maketitle

\begin{abstract}
We explore the possibility of accelerating the formal verification of classical programs with a quantum computer.

A common source of security flaws stems from the existence of common programming errors like use after free, null-pointer dereference, or division by zero.
To aid in the discovery of such errors, we try to verify that no such flaws exist.

In our approach, for some code snippet and undesired behavior, a SAT instance is generated, which is satisfiable precisely if the behavior is present in the code.
It is in turn converted to an optimization problem, that is solved on a quantum computer.
This approach holds the potential of an asymptotically polynomial speedup.

Minimal examples of common errors, like out-of-bounds and overflows, but also synthetic instances with special properties, specific number of solutions, or structure, are tested with different solvers and tried on a quantum device.

We use the near-standard Quantum Approximation Optimization Algorithm, an application of the Grover algorithm, and the Quantum Singular Value Transformation to find the optimal solution, and with it a satisfying assignment.
\end{abstract}

\section{Introduction}

Software testing represents a critical phase in the software development lifecycle, ensuring the reliability, security, and proper functioning of software systems.
Proper software verification and validation also provide a lot of potential for cost savings, especially when applied early in the software development process.
By identifying and correcting flaws before they can propagate through later stages of development or, worse, into deployed systems, organizations can avoid the exorbitant costs associated with post-deployment fixes, the potential harm to users and damage to the organization's reputation.

Formal verification methods apply mathematical models to verify the correctness of software with respect to a formal specification of its intended behavior.
A typical specification, for example for a banking transaction system, could be that the balance of an account should never become negative after a transaction.
There is a variety of techniques and tools available that build a suitable abstract model from the source code and specification of properties that need to be verified.
Symbolic execution engines, e.g. \autocite{Cadar2008}, often result in a Satisfiability Modulo Theory (SMT) problems \autocite{Barrett2011}, other approaches are based on reachability in the code's so-called Control Flow Graph (CFG) \autocite{Allen1970}, which allows a more abstract view of programs.

Being able to formally proof properties of code is a huge advantage, however, the application of formal verification and model checking is not without its challenges.
Rice's theorem \autocite{Rice1953}, asserts that if a property does not hold for all or no program, no algorithm will correctly decide if the property holds for all possible programs.
This theorem sets a theoretical limit on the capabilities of formal verification, implying that we must severely restrict ourselves concerning the possible properties we wish to certify.

Moreover, the computational demands of formal verification are significant.
Even if a verification problem is decidable, it is very often also NP-hard.
This can also be understood intuitively by considering the CFG and realizing that every decision node potentially doubles the amount of paths that need to be checked.
Here, quantum computing might help to solve these types of problems more efficiently in the future.

In the following paper, we evaluate different possible approaches and quantum algorithms towards this goal.
To that end, we introduce an end-to-end pipeline all the way from code snippets to the solution of the satisfiability problem corresponding to the desired verification task.

Our approach starts with a file of C-code and takes the following steps.
\begin{enumerate}
\item The checked properties are specified as flags for the converter.
\item A SAT formula, that is precisely satisfiable if a problematic input exists, is generated.
\item An optimization problem is generated from the resulting formula.
\item The optimal solution is found with the help of a quantum device.
\end{enumerate}
Different solvers are used and compared.

In summary, our contributions are as follows:
\begin{itemize}
\item A technique to transform a SAT instance to a QUBO with a guaranteed gap between possible values and its minimum depending on the satisfiability of the instance.
\item An end-to-end implementation of the full verification task from a C-file to the guarantee of a flaw or its likely absence.
\item The initial assessment of three different approaches to solving the resulting optimization problem with a quantum algorithm and comparison of some design choices for them.
\end{itemize}

\subsection{Structure}

We start in \prettyref{sec:Generating SAT} with some intuition as to how these formulae are constructed and how we generate them.
\prettyref{sec:Generating OP} explores the construction of the optimization problem from the formula.
As the last step, we see in \prettyref{sec:Solve}, how we solve the optimization problem with the use of a quantum computer.
With this, we present our performed experiments in \prettyref{sec:Tests} and compare how the used solvers and their variations perform on specific instances in \prettyref{sec:Results}.

\section{Related Work}

The exploitation of quantum computing in software verification is a new approach to harness quantum mechanics to improve the efficiency of verifying classical software systems.
To the best of our knowledge, the work in this paper represents the first practical attempt to apply quantum computing to accelerate the verification of classical software systems.

To date, the majority of related research has focused on problems such as quantum circuit verification \autocite{Ying2021} or software verification with classical computers, like model checking \autocite{Beyer2022}.
However, there has been a gap in the literature when it comes to applying quantum computing to classical software verification tasks.

With all its theoretical works, there is also an abundance of tools for different verification tasks \autocite{Davis2022}.

There are different approaches to software verification, that theorists strive to unify, e.g. \autocite{Riedmaier2020}.

To make verification possible for large systems, one approach is to divide the whole system into individual components \autocite{Malik2023}.
But even perfect protection from errors is worthless, if it is not applied in practice, may it be for the cost in time and resources or because only a specialized expert could use the techniques \autocite{Kaleeswaran2023}.

\section{Generating a SAT instance}\label{sec:Generating SAT}

In this section, we turn the question of whether some property holds for our code, into a logical formula.

\subsection{Introduction to CNF-SAT}

The Satisfiability Problem (SAT) asks for a given boolean formula if there exists an assignment of its variables that satisfies the formula.
We will specifically consider the Conjunctive Normal Form (CNF).

A formula in CNF is a conjunction of clauses, where a clause is a disjunction of literals and a literal is a variable or negated variable.
For \(0\leq l\leq 2^n\), it is of the form
\[
\bigwedge_{i=1}^l\bigvee_{j=1}^{k_i} L_{ij}.
\]
To also be in normal form, there are some additional constraints like no repeating variables in a clause an no clause being a sub-clause of another.

Every formula can be converted to this form with polynomial overhead, so we can assume the input to already be given in CNF.

\subsection{Expressing the Existence of Errors as a Formula}

By formally verifying software, we mean proving that some defined property holds for every input to the software.
Typical properties of interest are related to run time or general software errors, such as no buffer ever overflowing or a function output always being inside a valid range.
Such a property reduces the number of possibilities for a potential attacker to exploit the system and ensures that the program behaves as expected even for unaccounted inputs.

One approach is to construct the Control Flow Graph (CFG) of the program and encode the property as good and bad states.
How precisely the logical expression is obtained from the CFG is a problem of its own, but in simple cases it can be seen as extracting every path from the start to the bad states.
This explanation should give some intuition to the procedure, see \autocite{Clarke2018} and \autocite{Kroening2017} for a much more complete explanation.

Normally, the resulting formula will contain integers and could even be of higher order.
To handle these problems, we can make restricting assumptions to the environment.
Integers can be upper- and lower-bound, and higher-order operators can be bound with further assumptions about the possible models.
This means that our process will only find valid solutions, making it correct, but if no solution is found, one outside the assumed restrictions might still exist, so it is no longer complete.

For this work, the conversion is performed by existing tools, concretely CBMC \autocite{Clarke2004}, where the output is always a SAT instance.
CBMC can be configured to what errors it checks for and performs the needed modifications automatically.
Other tools, such as Seahorn \autocite{Gurfinkel2015}, were considered and some are considered to be used in future work.

The resulting SAT formula is parsed with PySMT \autocite{Gario2015} and optionally simplified by Z3 \autocite{Moura2008}, a symbolic logic as well as SAT solver.
This is not done for this work, which we will explain in \prettyref{sec:Tests}

\section{Generating the Optimization Problem}\label{sec:Generating OP}

The Quadratic Unconstrained Binary Optimization (QUBO) formulation is particularly fitting for quantum computing, because it naturally maps binary variables and objective functions into the language of qubits and quantum gates.

A QUBO is given by
\[
\argmin_{x\in\{0,1\}^n}x^TQx+c
\]
for \(Q\in\mathbb{R}^{n\times n}\) and \(c\in\mathbb{R}\).
This formulation nicely maps to the Ising model, see \autocite{Lucas2014} for details, that in turn can, for some approaches to hardware, be directly mapped to qubits.
This close relation to potential hardware made them very popular \autocite{Punnen2022}.

Note that this can be seen as a quadratic polynomial with the variables \(x_1,\ldots x_n\), where the linear terms are in the diagonal of \(Q\).

\subsection{Satisfiability as Optimum in a Quadratic Program}

To convert a logical formula into a QUBO, we have to map its logical structure to an arithmetic equivalent.
This has to be done carefully to keep all valid assignments, but also not introduce new ones.
Since we can restrict us to the CNF case, the conversion will be simplified as well.

We are searching for satisfying assignments, while the optimization problem has minimal assignments as its solution.
Therefore, the arithmetic expressions will be chosen to have their minima at precisely the satisfying assignments.
Our goal is to construct a QUBO, so quadratic polynomials will be used in our case.

We also make the choice to lower bound the polynomials by zero for values in \(\{0, 1\}\), which can be done by choosing the linear terms correctly.
By constructing the minima of all our polynomials to be at zero as well, we ensure that the minimum of the sum of multiple will also be precisely at zero.
This allows us to convert the clauses of our CNF formula to polynomials and simply take the sum over them to represent the full formula.

As another design choice, we choose our coefficients to be whole numbers.
This has the nice effect of enforcing a gap for the possible objective values, meaning the optimum reaches zero or is at least one.

For a clause with two literals, we get the conversion
\[a\lor b\to 1 -a -b +ab.\]
More literals would result in higher degree polynomials, if we apply this schema again.

To get around this issue, additional variables are introduced.
To reduce two literals (\(a, b\)) into one (\(r\)) the following polynomial can be used.
\[
    (1 - 2a - 2b)\cdot r + a + b + ab
\]
Adding it to the objective and replacing \(a\lor b\) by \(r\), reduces the literal count of the clause by one.
Since it has its minima precisely for \(r\iff a\lor b\), we get one to one correspondence of old and new minima.
This can be repeated, until only two literals remain.

Should a literal correspond to a negative variable (\(\neg x\)), one can use the same construction with \(x\) and simply replace \(x\) by \(1 - x\) afterwards.

This covers all needed constructions for CNF, but leaves the question of how many variables will be needed after the conversion.
Variables are only introduced for clauses that have more than two literals.
For a formula
\[
\bigwedge_{1\leq j\leq k}\bigvee_{1\leq i\leq n_j}L_{ij},
\]
this schema results in \(\sum_{j=1}^n \max\{0, n_j - 2\}\) extra variables.

\section{Finding the Optimum}\label{sec:Solve}

Different to usual optimization of looking for the optimal solution, we only care if the optimal value is zero or higher.
This precisely corresponds to the checked problem appearing for some input, turning the decision problem into an optimization one.
While it would technically be possible to gain such an input from an optimal solution, some steps for this are not trivial, and will be left for future work.
For this work, we only care if a flaw appears in the code or not.

We explore three approaches to search for an optimal solution.
\begin{description}
\item[VQA]
Our first approach are variational algorithms.
We use the Variational Quantum Eigensolver (VQE) \autocite{Tilly2021}, the Quantum Approximation Optimization Algorithm (QAOA) \autocite{Farhi2014} and Random Access Optimization (RAO) \autocite{Fuller2021}, provided by Qiskit as a subroutine.
They will be applied with different optimizers and ansätze.
\item[Grover]
Another way is to use Grover Amplification with the formula as oracle \autocite{Grover1996}.
Since Grover has a quadratic advantage in its regime, the hope is to find one here as well.
\item[Eigenvalue Filter]
Our last approach employs the Quantum Singular Value Transformation (QSVT), see \autocite{Tang2023} and \autocite{Gilyen2019}, together with a polynomial to filter a specific range of singular values \autocite{Lin2019}.
By only letting the states with value zero pass, only the solution space remains.
\end{description}
A comparison between them will be made in \prettyref{sec:Tests}.

\subsection{VQA}

VQE and QAOA are somewhat standard, but RAO will be given some motivation.
It is very similar to QAOA, but uses a Random Access Code (RAC) \autocite{Ambainis2009}.
It allows up to three variables to be encoded in one qubit, but one measurement can only extract one variable with a probability of about \(0.79\).
With RAO, we can potentially reduce the needed memory to a third, but need more than three times the number of executions to gain a solution \autocite{Fuller2021}.
It will also need a strategy for deciding what measurements to perform and how to interpret the results, called the rounding.

All three need an optimizer, while VQE and RAC also need an ansatz to be specified.
To decide on which to use, we decided on a set of each and performed a large trial to test all combination on our test cases with a simulated backend.
Based on these results, we choose candidates that converge fast, have a low number of iterations, and still find an optimal, or near optimal, solution.

Our picks for optimizers are COBYLA \autocite{Powell1994}, L\_BFGS\_B \autocite{Liu1989} and SPSA \autocite{Spall1998}.
For our Ansätze, we choose RealAmplitudes, EfficientSU2 (both part of qiskit) and PauliTwoDesign \autocite{Nakata2015}.

The last option is what rounding to use for RAO.
We use the "MagicRounding", since it performs much better than its alternative, "SemideterministicRounding".

\subsection{Grover}

For this approach, the formula is used to build a phase oracle, so the optimization problem is not constructed.

Since the number of solutions is needed to decide the number of repetitions of the Grover operator, we take the approach of testing every power of two with multiple shots. The resulting distribution is filtered for likely results.

For \(2^k\) solutions, a perfect amplification would result in a probability of \(2^{-k}\) to measure one specific solution.
We filter for solutions that have a likelihood of at least \(\frac{2}{3}\) of the expected one.

If none remain, we also add a control to the oracle to artificially double the search space, while keeping the number of solutions.
This is for the case that half, or more, of the possibilities are solutions.
This would mean in our setting, that more inputs than not have the undesired behavior, which should not be a state in which one tries to verify software.

\subsection{Eigenvalue Filter by QSVT}

Since the QSVT was shown to incorporate nearly all the major quantum algorithms, we wanted to use it for optimization as well.
The specific approach will be described in greater detail, because of the challenges that arise when actually implementing it.
We generally follow a procedure shown in \autocite{Martyn2021}.

QSVT allows one to apply a polynomial \(p\colon [-1, 1]\to [-1, 1]\) to the singular values of any matrix \(A\in\mathbb{R}^{n\times m}\) with \(\Vert A\Vert\leq 1\) and applies the resulting matrix to the current state, up to post-selection on an ancilla qubit.

As we showed at the end of \prettyref{sec:Generating OP}, it is possible to enforce all satisfying assignments to take the value zero, while all unsatisfying ones have at least a value of one.
We will therefore encode our QUBO as the matrix \(\delta A\), for \(\delta := \Vert A\Vert\), and apply an approximation of the threshold function
\[
[0,1]\to [0,1],\, x\mapsto \begin{dcases*}
1 & if \(x=0\)\\[1ex]
0 & if \(|x|\geq\frac{\delta}{2}\)
\end{dcases*}
\]
to only leave solutions after post-selection.
This means that an unsatisfiable formula leaves no results, because they would all be removed by post-selection.

The general architecture are alternating layers of controlled rotations that depend on \(p\), and operators encoding \(A\).
After \(\deg(p)+1\) layers, overall, the operation \(p(A)\) is performed, up to the post-selection of an additional qubit.
For details see \autocite{Tang2023} or \autocite{Gilyen2019}.

We start by constructing a circuit that encodes our QUBO (\(x^TQx+c\)).
The QUBO is first formulated in the Ising Model \autocite{Lucas2014}.
This results in a Hermitian matrix \(H\) with
\[
H\ket{x}=(x^TQx+c)\ket{x}
\]
for all base states \(\ket{x}\).
This \(H\) is the matrix we wish to encode as the matrix \(A\), but we still have to find an appropriate scale \(\delta\).

We can find an upper bound to the highest singular value by bounding every clause and summing over them, which we call \(M\).
This allows us to scale the singular values into the valid region \([0, 1]\), by the factor \(\delta:=M\).

If we set \(A:=\delta H\), we can use a very general approach to define a unitary.
\[U':=\begin{psmallmatrix}
A & \sqrt{I - AA^\dag}\\
\sqrt{I - A^\dag A} & -A^\dag\\
\end{psmallmatrix}\]
By padding this matrix with ones on the diagonal and otherwise zeros to a dimension that is a power of two, we get \(U\), from which a quantum circuit will be constructed that can be used as part of the QSVT.

Next, we need to decide what polynomial to apply.
The perfect function with maximum distance from important values to a discontinuity is
\[
[-1,1]\to [-1,1],\, x\mapsto \begin{dcases*}
1 & if \(|x|<\frac{\delta}{2}\)\\[1ex]
0 & if \(|x|\geq\frac{\delta}{2}\)
\end{dcases*}.
\]
But finding a good approximation is not easy.

We will use the polynomials from \autocite{Lin2019}, since they fit our needs nicely.
They are defined by
\[
F_{2d,\delta}(x):=\frac{
    T_{d}\left(
        2\frac{
            x^2 - \delta^2
        }{
            1 - \delta^2
        } - 1
    \right)
}{
    T_{d}\left(
        \frac{
            -2\delta^2
        }{
            1 - \delta^2
        } - 1
    \right)
}
\]
where \(2d\) is the degree of the resulting polynomial, \(\delta\) is the gap and \(T_d(x)=\cos\left(d\arccos(x)\right)\) is the Chebyshev polynomial of the first kind.

\section{Test Cases\label{sec:Tests}}

\prettyref{tab:Examples} contains an overview of all tested error types and logical formulae.
\begin{table}[tbh]
    \centering
    \caption{Used examples of software errors and logical expressions.}
    \label{tab:Examples}
    \begin{tabular}{l l}
        \toprule
        \textbf{Name} & \textbf{Test Case}\\
        \midrule\midrule
        Out of Bounds & 
\begin{lstlisting}[language=C, tabsize=4]
int buf[1];
buf[1] = 0;
\end{lstlisting}\\
        \midrule
        Invalid Conversion & 
\begin{lstlisting}[language=C, tabsize=4]
short s = 1024;
char c = (char) s;
\end{lstlisting}\\
        \midrule
        Division by Zero & 
\begin{lstlisting}[language=C, tabsize=4]
float i = 1 / 0;
\end{lstlisting}\\
        \midrule
        Memory Leak & 
\begin{lstlisting}[language=C, tabsize=4]
void *p = malloc(1);
p = 0;
\end{lstlisting}\\
        \midrule
        Not a Number & 
\begin{lstlisting}[language=C, tabsize=4]
float i = 0.0 / 0.0;
\end{lstlisting}\\
        \midrule
        Overflow & 
\begin{lstlisting}[language=C, tabsize=4]
int k = 0x7FFFFFFF;
k += 1;
\end{lstlisting}\\
        \midrule
        Null-Pointer Dereferencing & 
\begin{lstlisting}[language=C, tabsize=4]
int *p = 0;
int i = *p;
\end{lstlisting}\\
        \midrule
        Addition & \(a+b=2c + d\)\\
        \midrule
        Program Flow & \((a=b=c)\land (d+e+f>1)\)\\
        \midrule
        Indicator & \(2a+b > 2c+d\)\\
        \midrule
        OR & \(\bigwedge_{j=1}^nx_j\)\\
        \midrule
        XOR & \(\bigoplus_{j=1}^nx_j\)\\
        \midrule
        Unique & \(\sum_{j=0}^52^jx_j\in\{42\}\)\\
        \midrule
        Semi-Unique & \(\sum_{j=0}^72^jx_j\in\{42, 69\}\)\\
        \midrule
        Two-Solutions & \(\sum_{j=0}^{13}2^jx_j\in\{15, 240\}\)\\
        \midrule
        Two-Solutions-Overlap & \(\sum_{j=0}^72^jx_j\in\{85, 204\}\)\\
        \midrule
        Three-Solutions & \(\sum_{j=0}^72^jx_j\in\{42, 101, 205\}\)\\
        \bottomrule
    \end{tabular}
\end{table}
While they look different, many turn out to lead to the same logical expression, resulting in a reduction of solved optimization problems.
More complex examples were tried, but even slightly larger code snippets can very easily generate huge SAT instances with hundreds or even thousands of variables.
A simple malloc with dropped pointer results in 341 variables, a simple gcd implementation with maximal iteration depth four needs 175701 variables, and at depth six 704949 variables.
Especially loops in the CFG result in too many variables to be handled by QC.
It is also impossible to tell how large the SAT for a code snippet becomes, without performing the conversion.

To allow us to have optimization problems with finer control over their sizes, some additional synthetic SAT instances were created.
They are inspired by our minimal software errors, arithmetic, and program flow, or were chosen because of special logical properties.
Creating them synthetically, allows us to enforce a specific number of variables.

Note that the number of variables counts the binary variables of the logical formula, not the variables in the corresponding optimization problem.
This number will likely be higher and corresponds to the needed number of qubits for most approaches.

\section{Results\label{sec:Results}}

We had access to a 27 qubit IBM Quantum System One ibmq\_ehningen with Falcon r5.11 architecture, and later the ibmq\_nazca system with 127 qubits with Eagle r3 processor.
No error mitigation techniques besides the defaults of the sampler and estimator primitives were used.

\subsection{VQA}

The experiments were performed in two parts
\begin{enumerate}
    \item All test cases were executed using COBYLA and RealAmplitudes on the Ehningen system.
    \item All optimizers with all Ansätze were used to solve the ADD test case on the Nazca backend.
\end{enumerate}
This was done because execution time was limited.

The first part was performed on the Ehningen system.
All test cases were solved by QAOA with COBYLA.
While a solution was found for all of them, especially the largest instances took a long time to converge.
We were able to conclude that larger problems will need more iterations to still find the solution from our experiments.

The second part is meant to compare how a different solver and ansatz change the convergence rate and was performed with tighter tracking of the convergence.
We wanted to choose a test case that is not large (to ensure completion within our calculation window), has not too many solutions (to fit our motivation), and is at least algorithmically motivated.

\begin{figure}[htb]
    \centering
    \includegraphics[width=\linewidth]{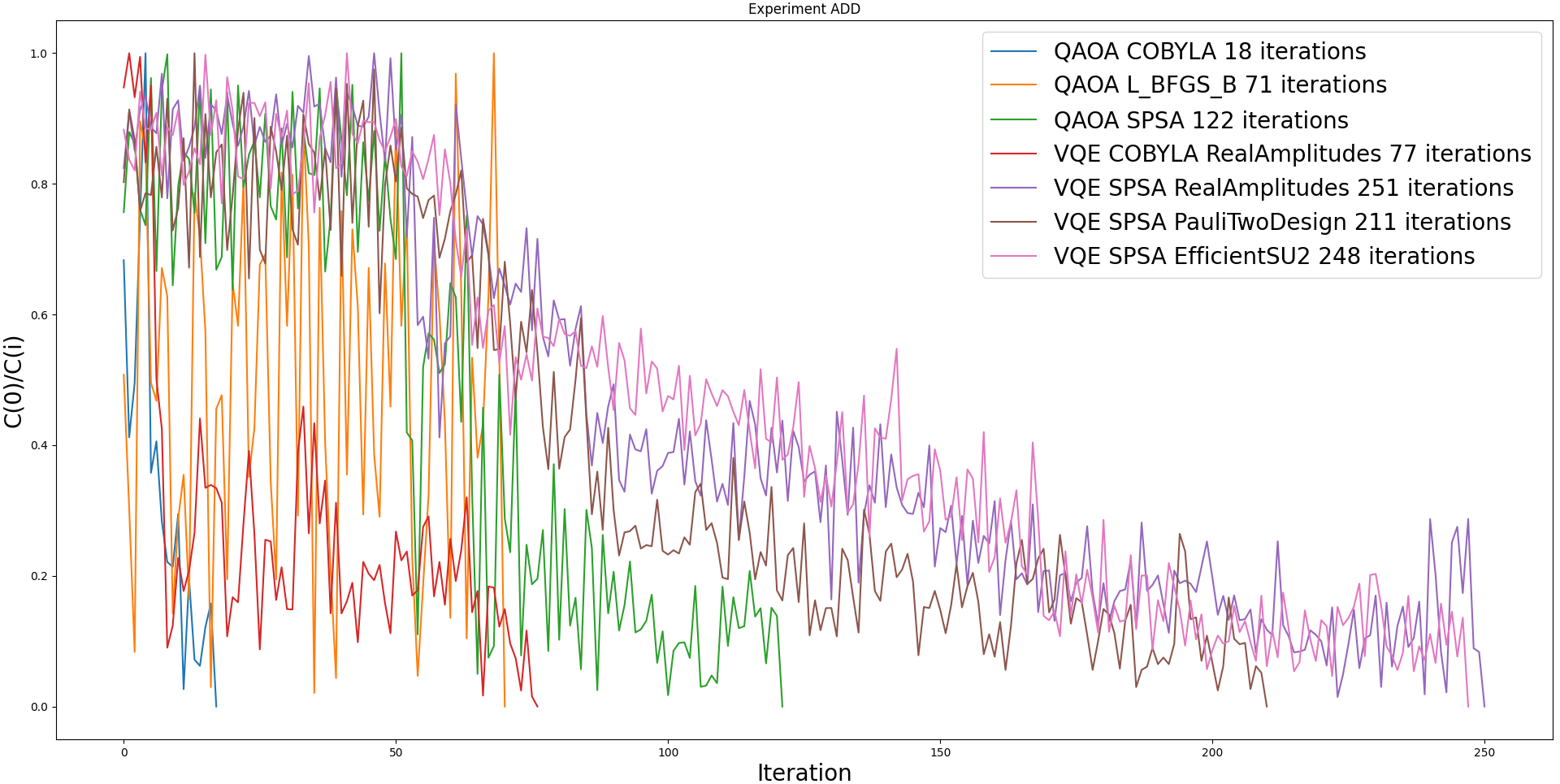}
    \caption{Convergence of QAOA and VQE for different solver and ansatz.}
    \label{fig:QAOA Result}
\end{figure}

The convergence of the runs to the first minimum for VQE and QAOA can be found in \prettyref{fig:QAOA Result}.
The plot depicts how the current value relates to the starting value and the optimum, normalized to the interval \([0,1]\).
Not all combinations appear in the figures, because several failures were noted during the experiments.
The VQE trials using L\_BFGS\_B and COBYLA, except for RealAmplitudes, did not finish.
At some time during the run, the communication to the backend failed.

We see that QAOA converged significantly faster than the others and the best fastest with the COBYLA optimizer.
We can also see that the SPSA optimizer converges much slower for our test case.
The COBYLA optimizer seems to perform significantly better than the others.

\begin{figure}[htb]
    \centering
    \includegraphics[width=\linewidth]{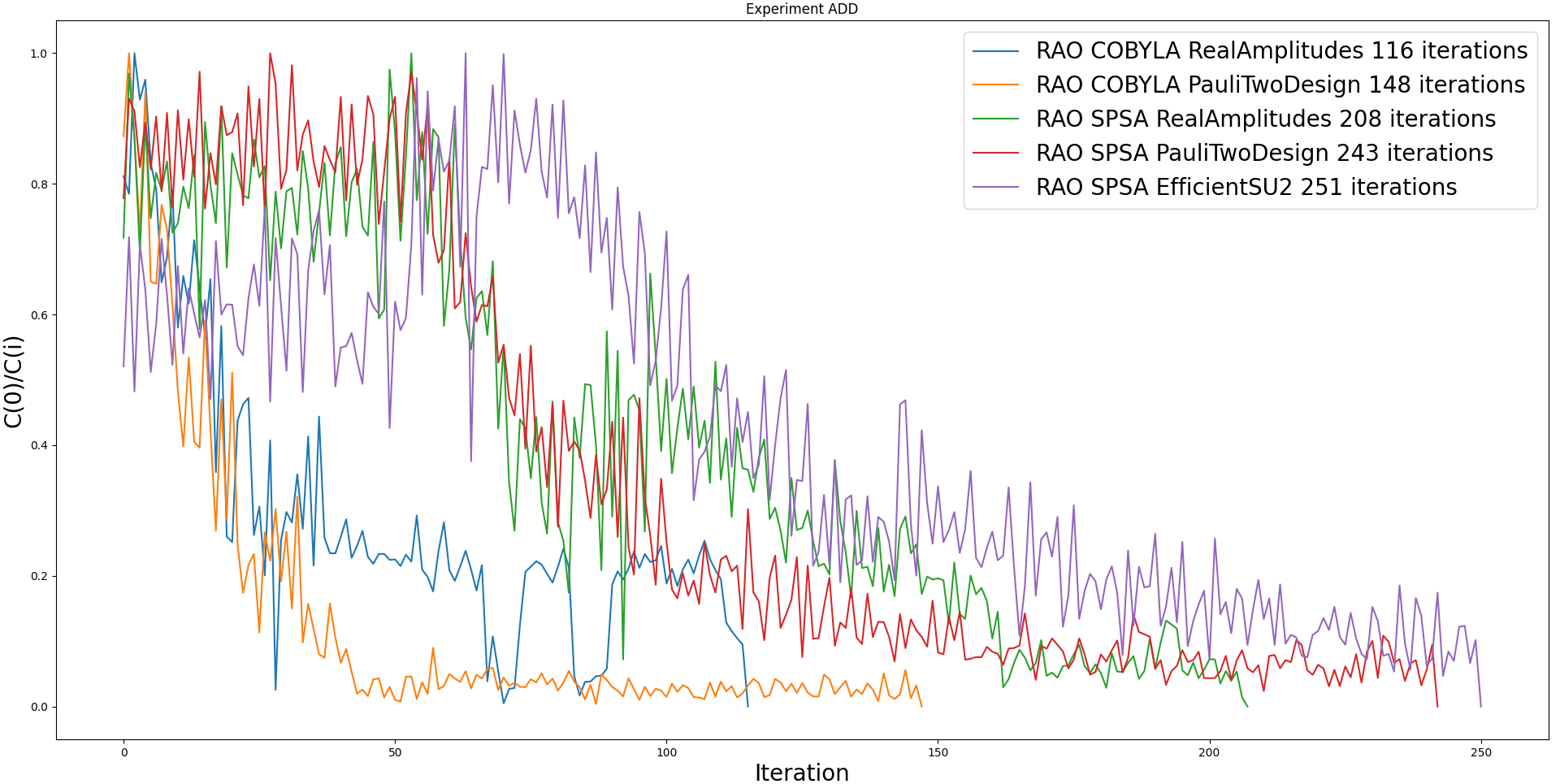}
    \caption{Convergence of RAO for different solver and ansatz.}
    \label{fig:RAC Result}
\end{figure}

The convergence of RAO to the first minimum is illustrated in \prettyref{fig:RAC Result}.
The RAO tests with L\_BFGS\_B and COBYLA using EfficientSU2 also experienced failures, attributed to difficulties in minimizing the encoded problem.

COBYLA is again significantly faster at a minimum.
Here we can also note that the RealAmplitude ansatz converges the fastest.

We can note that RAO needs more iterations, but is definitely competitive to VQE, while compressing \(2.8\) variables into one qubit.
This is especially interesting for larger problems, because they could fit for RAO, while being too large for the other approaches.

\subsection{Grover}

All instances of the tests were executed on the Nazca backend.
While the experiments performed exceptionally well on the simulator, the results obtained on actual hardware were nearly nonexistent.

Notably, no test returned a correct result with high enough probability, except for the XOR(2) instance, which achieved one correct output, but not the other.
This result could be attributed to chance, as half of all assignments are expected to yield hits.
A closer examination of the full distributions indicated that the outcomes were only marginally better than random results.

This outcome is not particularly surprising, considering that the noise present in the quantum hardware significantly undermines any advantages the Grover algorithm may provide.
Given that Grover's algorithm has an exponential runtime, it is inherently prone to high error amplification, further complicating the extraction of meaningful results from the hardware tests.

\subsection{Eigenvalue Filter by QSVT\label{sec:ExQSVT}}

The described technique was successfully used for the smaller instances on a simulated backend, but we were not able to run the experiments on hardware.
Multiple problems arose, but specifically the construction of a block encoding of the matrix turned out to be extremely challenging.
While it was possible to construct the problems for slightly larger instances, solving them, turned out to need more memory then was available at the moment.

For our block encoding a potentially arbitrary unitary matrix must be transpiled to a quantum circuit using only the supported basis gates.
While this is technically possible, the transpilation could not be performed even for 8 qubits on a 16 GB laptop computer due to memory exhaustion.
The simulator simply used the matrix, but execution on hardware is impossible.
While there are approaches for better encodings, they will be left for future work.

For an overview of our performed experiments, see \prettyref{tab:Results}.
\begin{table}[tbh]
    \centering
    \caption{Results of Experiments, grouped by resulting optimization problem.
    Gap and degree are \(delta\) and \(2d\) for \(F_{2d,\delta}\), and rate is the success probability of one shot.}
    \label{tab:Results}
    \begin{tabular}{l c c c c}
        \toprule
        \textbf{Instance} & \textbf{\#Variables} & \textbf{Gap} & \textbf{Rate} & \textbf{Degree}\\
        \midrule\midrule
        Addition & 8 & 1/9 &  & 51\\
        \midrule
        Flow, Indicator & 4 & 1/5 & 0.36 & 21\\
        \midrule
        OR(3) & 4 & 1/4 & 0.43 & 17\\
        \midrule
        XOR(2) & 6 & 1/4 & 0.021 & 17\\
        \midrule
        XOR(3) & 7 & 1/10  &  & 53\\
        \midrule
        Unique & 6 & 1/6 & 0.011 & 30\\
        \bottomrule
    \end{tabular}
\end{table}
Note that the provided gap is the best possible instead of the mentioned estimate and many cases had to be dropped for this approach, due to being too large.

Degrees were chosen empirically, based on the following metric and high enough to only sample a miniscule amount of non-solutions after post-selection.

\begin{figure}[htb]
    \centering
    \includegraphics[width=\linewidth]{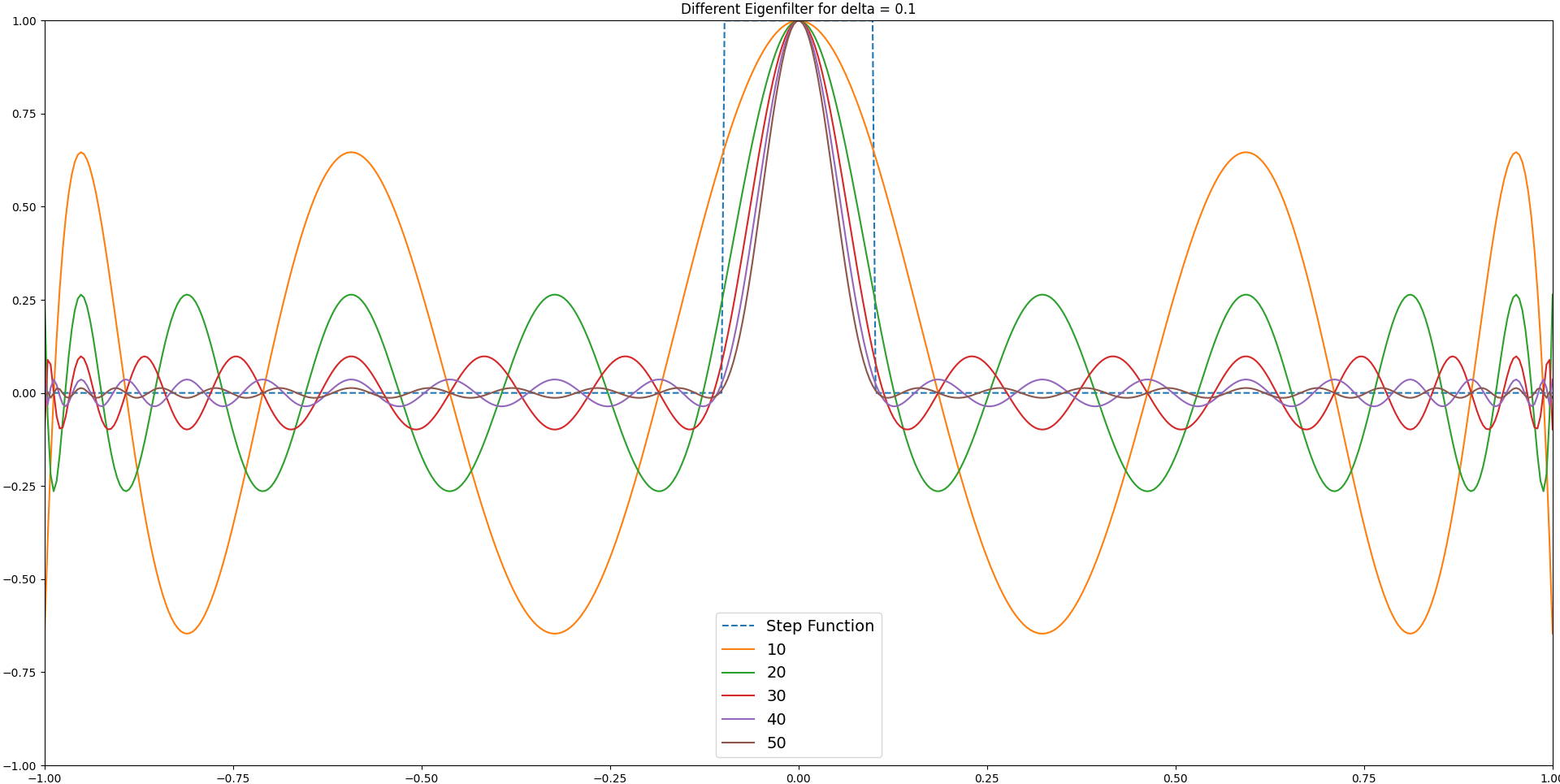}
    \caption{The used filters \(F_{d,\delta}\) with different degree at \(\delta=\frac{1}{10}\).}
    \label{fig:Filter tight}
\end{figure}

To see what impact the degree has on the filter function, see \prettyref{fig:Filter tight} for \(\delta=\frac{1}{10}\).

We see that a high degree is needed to amplify solutions, while also suppressing non-solutions.
Smaller gaps would need even higher degree to be useful.
To visualize how usable the approximation is, we introduce the following measure for a function \(f\) and gap \(\delta\).
\[
\mu_f(\delta):=\frac{
    \vert f(0)\vert^2
}{
    \max_{j=1}^{\lfloor\delta^{-1}\rfloor}
    \vert f(j\delta)\vert^2
}
\]
It measures how much a solution is amplified, compared to any non-solution.
This uses the fact that for our construction, values can only be multiples of the gap.

We want an amplification to a probability of success to at least a half, even in the worst case, which for \(n\) qubits has a search space of size \(2^n\) and only one solution, so
\[\mu\geq 2^{n-1}\iff 1 +\log_2(\mu)\geq n.\]

\begin{figure}[htb]
    \centering
    \includegraphics[width=\linewidth]{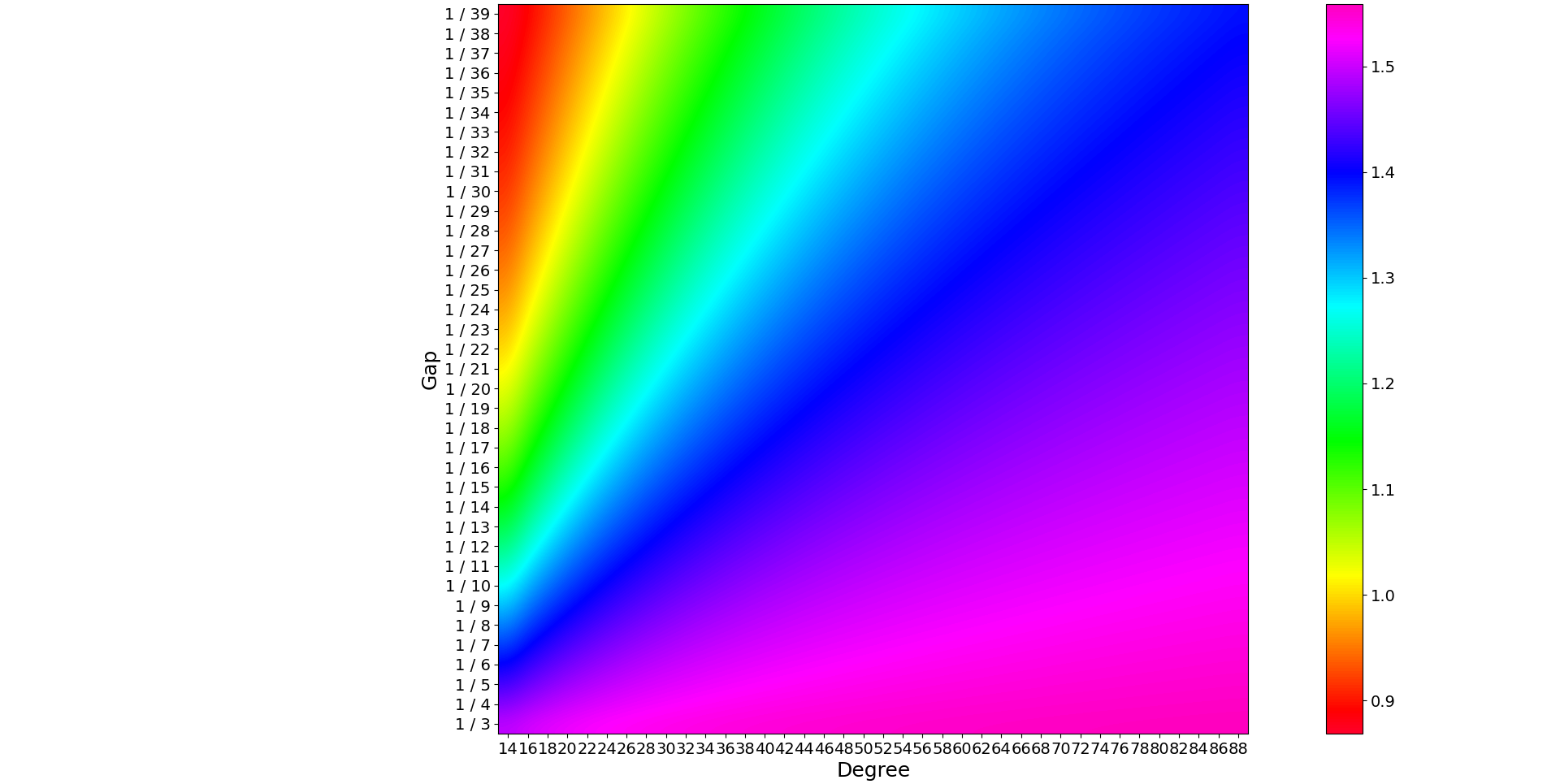}
    \caption{Heatmap of the quality of the approximation for the gap \(\delta\) given by the y-axis with degree \(d\) given by the x-axis, according to the explained measure \(\mu_{F_{d,\delta}}(\delta)\).}
    \label{fig:Heatmap}
\end{figure}

To visualize what number of qubits, and therefore problem sizes, can be handled, we plot \(\arctan(1+\log_2(\mu_{F_{d,\delta}}(\delta)))\) in \prettyref{fig:Heatmap}.
To enforce an upper bound, while not changing small values, the \(\arctan\) is used as well.

A more complex problem has a smaller gap, which results in the need of a larger degree to amplify the space of solutions significantly enough.

Another problem is the need for post-selection.
Even for perfect amplification, many shots are removed, since they measure the wrong subspace.
This is because the filtered matrix acts as a projector, which is enforced by the post-selection.

If the problem has only one solution and apply a perfect filter, we might still have a low probability of getting the desired result.
Preparing the superposition of all possible solutions leaves only a small probability for every individual value.
After the filter, this likelihood does not increase, only the non-solutions are now removed by post-selection.

\section{Conclusion}

As expected, the process is possible and always returns the correct result, even in a noisy setting.
However, the returned result is correct, but not complete, so only a probabilistic certification is possible.
It can only guarantee the absence of the error, precisely as much as that the actual minimum is found.

All solvers face similar challenges in slightly different ways.
They all need significantly longer, once the gap gets smaller.
The Grover approach did stick out, since it solved one trivial problem on actual hardware, but failed for all others.
Whereas the QSVT application was limited by the precision of the classical calculations and the implementation of the block encoding.

From the variational solvers, QAOA had the fastest convergence, especially with the COBYLA optimizer, but with smaller gaps it starts to loose precision or, equivalently, needs more iterations to converge.
RAO did surprisingly well, considering its significantly smaller memory need.
It took about 2-3 times as many iterations to find the optimal solution, excluding the best QAOA run, while using about a third of the qubits.
This can allow problems to be solved, that would otherwise be too large for the hardware.

The number of needed iterations clearly depends on the size of the problem, but is also depends on other factors.
There even seems to be a relation between the number of iterations to the gap in the QSVT approach, the smaller the gap, the more iterations are needed.
This might even generalize to the annealing framework, where the annealer would need a longer relaxation time.

Grover performed as expected.
Small instances perform very well on a simulated backend, but it still has an exponential runtime, which makes larger instances impossible to approach.
On hardware, nearly all amplification of solutions is overshadowed by the accumulated noise.

The major bottleneck to using the QSVT on hardware is the block encoding and needed degree.
The block encoding has to be applied equal to the degree plus one, together with some phase operations.
Should a technique arise to perform this step in a hardware-friendly way, it could turn into a useful approach.
its usefulness also depends on another problem, the failure probability of post-selection.
If it turns out to scale too strong with the problem size or gap, then the overall likelihood of success is no better than guessing.

We have shown that moving the verification step to a quantum computer is possible.
From this we can conclude, that an improvement to optimization in QC over classical computing, would also be possible to be reapplied to software verification.
The practicality at the moment is very limited.
While the numerical problems can be dealt with, the noise on actual hardware is a major issue.
Should error corrected Quantum Computing arise, a performance improvement would be possible as well.

\subsection{Future Work}

There are many research directions for possible future work and some high-level points are now listed.

While we tried to make reasonable choices for our encoding, others might have advantages we did not expect, like fitting well on specific hardware architectures or resulting in fewer variables for some interesting class of instances.
Testing different hardware and simulators with different encodings could be a fruitful approach.

One could use weights in the logical formulae to enforce specific constraints to hold.

Since QUBO is nearly a native formulation for the annealing paradigm, the use of a quantum annealer and how well different encodings of the satisfiability problem fit on the annealer architecture could show what problems are well suited for QC in general.

Finally, better block encoding techniques might allow to run our QSVT application on quantum hardware.

\printbibliography

\end{document}